\documentclass[8pt,a4paper,twocolumn]{article}
\usepackage[margin=20mm]{geometry}
\usepackage{cite}
\usepackage{amsmath,amssymb,amsfonts}
\usepackage[ruled]{algorithm2e} 
\usepackage{graphicx}
\usepackage{textcomp}
\usepackage{xcolor}
\def\BibTeX{{\rm B\kern-.05em{\sc i\kern-.025em b}\kern-.08em
　　T\kern-.1667em\lower.7ex\hbox{E}\kern-.125emX}}
\begin{document}

\title{\Huge Pairs-trading System using Quantum-inspired Combinatorial Optimization Accelerator for Optimal Path Search in Market Graphs}

\author{
Kosuke Tatsumura$^{\ast}$, Ryo Hidaka, Jun Nakayama, \\
Tomoya Kashimata, and Masaya Yamasaki \\
\small Corporate Research and Development Center, Toshiba Corporation, Japan\\
$^{\ast}$\small Corresponding author: Kosuke Tatsumura (e-mail: kosuke.tatsumura@toshiba.co.jp)
}
\date{}

\maketitle

\begin{abstract}
Pairs-trading is a trading strategy that involves matching a long position with a short position in two stocks aiming at market-neutral profits. While a typical pairs-trading system monitors the prices of two statistically correlated stocks for detecting a temporary divergence, monitoring and analyzing the prices of more stocks would potentially lead to finding more trading opportunities. Here we report a stock pairs-trading system that finds trading opportunities for any two stocks in an $N$-stock universe using a combinatorial optimization accelerator based on a quantum-inspired algorithm called simulated bifurcation. The trading opportunities are detected through solving an optimal path search problem in an $N$-node directed graph with edge weights corresponding to the products of instantaneous price differences and statistical correlation factors between two stocks. The accelerator is one of Ising machines and operates consecutively to find multiple opportunities in a market situation with avoiding duplicate detections by a tabu search technique. It has been demonstrated in the Tokyo Stock Exchange that the FPGA (field-programmable gate array)-based trading system has a sufficiently low latency (33 $\mu$s for $N$=15 or 210 pairs) to execute the pairs-trading strategy based on optimal path search in market graphs.
\end{abstract}

\section{Introduction}\label{sec:introduction}
A financial market with high efficiency and high liquidity is where investors can execute high-volume trading at fair values, at any time without significantly impacting the market prices. The concept of arbitrage is defined in Ref.~\cite{Sharpe90} as the simultaneous purchase and sale of the same, or essentially similar, security in two different markets for advantageously different prices. Arbitrage opportunities can arise as a result of demand shocks and arbitragers bring temporarily deviated prices (hereafter, mispricing) to fundamental (fair) values. Arbitrage enforces the law of one price and thereby improves the efficiency of financial markets~\cite{shleifer97}. Recent studies~\cite{gromb10, rosch21} have also shown that arbitrage provides liquidity.

Pairs-trading strategy is categorized as a statistical arbitrage and profits from temporary mispricing of statistically correlated stocks~\cite{gatev06}. The strategy monitors the performance of two historically correlated stocks for detecting the moment when one stock relatively moves up while the other relatively moves down (possibly temporarily), and at that moment simultaneously takes a short (selling) position of the outperforming stock and a long (buying) position of the underperforming one with each position having the almost same amount of transaction, betting that the spread between the two would eventually converge. The strategy is market-neutral, i.e., adaptable to various market conditions (uptrend, downtrend, or sideways) by keeping the net exposure low.

Various variants of pairs-trading that differ in how to identify comoving stocks and how to decide the timing of position opening have been proposed and summarized in Ref.~\cite{krauss17}, involving distance approach, cointegration approach, time-series approach, stochastic control approach and other approaches (including machine learning approaches like recent one using long short-term memory networks~\cite{flori21}). Those, not necessarily mutually exclusive, can contribute to improving the market efficiency and liquidity by detecting the different trading opportunities (occurrences of mispricing).

To analyze the collective structure of a stock market, market graphs have been proposed and utilized~\cite{butenko03,boginski04,marzec16}, where the nodes correspond to the stocks and each edge (or edge weight) between two nodes represents the relationship of the two stocks defined based on correlation factors~\cite{butenko03,boginski04} or more generalized risk-measures~\cite{marzec16}. Graph analysis methods such as partitioning, clustering, coloring, and path search may give insights into the collective structures/behaviors of the stocks. Many of those methods are formulated as combinatorial (or discrete) optimization problems and belong to the nondeterministic polynomial time (NP)-hard class in computational complexity theory~\cite{lucas14}. 

Ising machines are hardware devices that solve the ground (energy minimum)-state search problems of Ising spin models and can be of use for quickly obtaining the optimal (exact) or near-optimal solutions of NP-hard combinatorial optimization problems~\cite{sbm1, FPL19, sbm2, NatEle, kanao23, johnson11,king23, honjo21, pierangeli19, cai20, aadit22, moy22, sharma22, takemoto19, kawamura23, matsubara20, waidyasooriya21, okuyama19}. The Ising problem belongs to the NP-hard class~\cite{lucas14, barahona82}; a variety of notoriously hard problems including many graph analysis methods can be represented in the form of the Ising problem~\cite{lucas14}.

The Ising machine can be applied to automated trading systems~\cite{yoo23,fil20,huang19,denholm15,leber11} including ones executing pairs-trading and may enable the detection of trading opportunities based on the computationally-hard analysis of market graphs within the lifetime of the opportunities determined by the activities of other trading entities. Automated trading systems become increasingly important in financial markets~\cite{malceniece23,brogaard14} and the trading strategy enabled with emerging computing methodologies would complement the functionality of the market or contribute to mitigating the herding behaviors in financial markets~\cite{spyrou13}. The trading systems utilizing Ising machines as in~\cite{ISCAS20} have been, however, not extensively studied. Furthermore, the execution capability of such a trading system in terms of response latency needs to be validated in the actual market.

Here we propose a pairs-trading strategy based on an optimal path analysis in market graphs and show through real-time trading that the strategy is executable with an automated pairs-trading system using an embedded Ising machine for the optimal path search.

The market graph for $N$ tradable stocks (an $N$-stock universe) is an $N$-node fully-connected directed graph with edge weights corresponding to the products of instantaneous price differences and statistical correlation factors between two stocks. The trading opportunities (temporary mispricing of statistically correlated pairs) are detected by an optimal path analysis (a sort of collective evaluation) of the $N$-node market graph. As the embeddable Ising machine, we use a combinatorial optimization accelerator based on a quantum-inspired algorithm called simulated bifurcation (SB)~\cite{sbm1, FPL19, sbm2, NatEle, kanao23}. The algorithm of SB, derived through classicizing a quantum-mechanical Hamiltonian describing a quantum adiabatic optimization method~\cite{qbm}, is highly parallelizable and thus can be accelerated with parallel processors such as FPGAs (field-programmable gate arrays)~\cite{FPL19}. An SB machine (SBM) customized for the strategy operates consecutively to find multiple trading opportunities in an instantaneous market situation with avoiding duplicate detections by a tabu search technique. To examine the execution capability of the system, we compare the real-time transaction records of the system in the Tokyo Stock Exchange (TSE) with a backcast simulation of the strategy assuming the orders issued are necessarily filled.

The rest of the paper is organized as follows. In Sec.~\ref{sec_strategy} (trading strategy), we describe the proposed strategy and formulate the optimal path search in the form of quadratic unconstrained binary optimization (QUBO) mathematically equivalent to the Ising problem. Sec.~\ref{Sec_System} (system) describes the architecture of the system and its implementation details. Sec.~\ref{sec_expt} (experiment) describes the transaction records in the TSE and the execution capability of the system.

\section{Trading strategy}\label{sec_strategy}
\subsection{Path search-based pairs-trading}
The proposed strategy determines open pairs (a pair of long and short positions in two stocks to be taken) by an optimal path analysis of an $N$-node market graph representing a relative relationship in the prices of $N$ stocks. The evaluation of a pair is based on not only the direct path but also any bypass paths. Multiple pairs can be chosen in an instantaneous market situation.

\begin{figure}[b]
\centering
\includegraphics[width=8.3 cm]{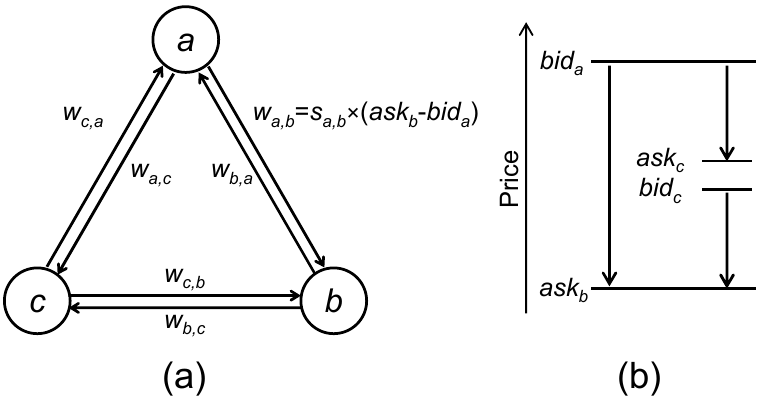}
\caption{(a) Market graph for an $N$-stock universe ($N=3$). (b) A relationship of $bid$ and $ask$ values regarding with the direct path ($a\rightarrow b$) and bypass path ($a \rightarrow c \rightarrow b$) evaluations of pair (a, b).}
\label{Fig_strategy}
\end{figure}

The market graph for an $N$-stock universe (Fig.~\ref{Fig_strategy}a) is a directed graph in which an edge ($i$,~$j$) corresponds to a trading pair that takes a short position of $i$th stock and a long position of $j$th stock and is distinguished from  the edge ($j$,~$i$).  The weight $w_{i,j}$ of an edge ($i$,~$j$) is defined by
\begin{equation}\label{Eq_wij}
w_{i,j}=s_{i,j}\times(ask_{j}-bid_{i})
\end{equation}
where $s_{i,j}$, $ask_{j}$, and $bid_{i}$ are, respectively, the similarity factor between $i$th and $j$th stocks, the best ask for $j$th stock, and the best bid for $i$th stock. $ask$ and $bid$ are normalized by the base price on the day. $s_{i,j}$ is based on the average value for the last five business days of the dynamic time warping (DTW) distance~\cite{sakoe78} of the price sequences (per day) of $i$th and $j$th stocks and is normalized to be in $[0,1]$. When the buying price of a long position ($ask_{j}$) is relatively lower than the selling price of a short position ($bid_{i}$) in the two stocks with a large similarity ($s_{i,j}$), $w_{i,j}$ is negative and its absolute value is large.

In the market graph, two nodes connected by the minimum-weight one-way directed path are considered to correspond to the best trading opportunity. A pair of nodes can be selected based on a bypass path rather than the direct path. In the case of Fig.~\ref{Fig_strategy}, the pair ($a$,~$b$) is evaluated for both the direct path ($a\rightarrow b$) and the bypass path ($a \rightarrow c \rightarrow b$). The bypass path corresponds to concurrently taking the pair ($a$,~$c$) and pair ($c$,~$b$) positions, leaving the pair ($a$,~$b$) position as a result of the cancellation of buying and selling the stock $c$ (the direct and bypass paths correspond to the same open pair). If not considering the similarity factors, the sum of $w_{a,c}$ and $w_{c,b}$ (bypass) is always higher than $w_{a,b}$ (direct) by the bid-ask spread of the stock $c$ (transit nodes on the bypass) (see Fig.~\ref{Fig_strategy}b). However, considering the similarity factors, the sum of $w_{a,c}$ and $w_{c,b}$ can be lower than $w_{a,b}$. In this case, the evaluation of pair ($a$,~$b$) is represented by the sum of the weights on the bypass path. This bypass evaluation (or collective evaluation) partially complements the incompleteness of the representation of similarity coming from characterizing time series data as a scalar value and prevents us from missing the trading opportunity. The evaluation value (weight sum) of a pair selected by the optimal path analysis is compared with a threshold for determining the opening of the pair.

The number of lots per order for a stock ($L_{i}$) is determined to make the amount of transaction ($A_\mathrm{trans}$) common for all tradable stocks by rounding with considering the minimum tradable shares per order (a lot) of the stock ($S_{i}^\mathrm{min}$) and the base price on the day ($p_{i}^{b}$); $L_{i}=round(A_\mathrm{trans}/S_{i}^\mathrm{min} p_{i}^{b})$. The number of intraday positions is controlled to be within a maximum number ($P_\mathrm{max}$) and all positions are closed (unwind) before the close of the day. Duplicate pair positions are not allowed. When the pair ($a$,~$b$) has been ordered (opened), another order of the same pair ($a$,~$b$) has been forbidden, but other pairs including ($a$,~$c$) and ($c$,~$b$) are orderable and the edge ($a$,~$b$) is passable for bypass evaluation.

Consider a subgroup of stocks (for an example, $a$, $b$, and $c$) that are correlated one another. If the price of one in the subgroup (assume $a$ in the example) deviates largely (drops in the example) while the prices of the remaining ones do not deviate, multiple pairs related to the deviating one [pairs ($b$,~$a$) and ($c$,~$a$) in the example] are highly evaluated at the moment and, as well as the best pair [pair ($b$,~$a$) in the example], the second-best pair [pair ($c$,~$a$) in the example] can be worth betting (can have an evaluation value beyond the threshold). To our backcast simulation (see Sec.~\ref{sec_expt}), a temporary price deviation of one stock in the cross-correlated subgroup gives good trading opportunities. For finding multiple opportunities in a market situation, the optimal path analysis is repeated. We need a sort of tabu search technique to avoid repeatedly finding the solution that has been found.

\subsection{Formulation}\label{subsec_formulation}
The problem to find a pair of two nodes connected by the minimum-weight directed path (direct or bypass) from any two nodes in the $N$-node market graph is formulated in the form of the QUBO. A tabu search technique using a tabu list ($T_{i,j}$) is implemented in the formulation.

\begin{figure}[t]
\centering
\includegraphics[width=8.3 cm]{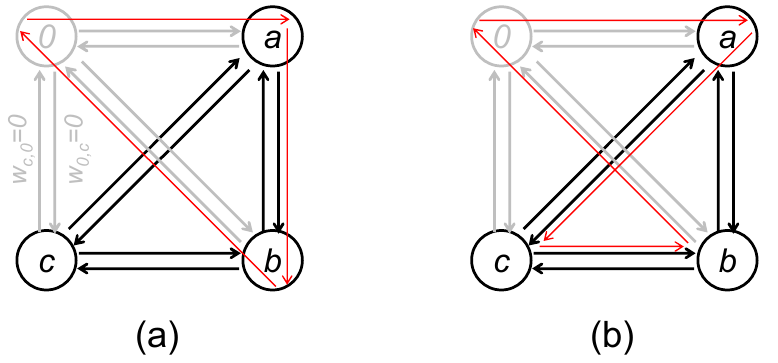}
\caption{(a) Market graph with the dummy node  ($i=0$) for an $N$-stock universe ($N=3$). (a) A cyclic path ($0\rightarrow a\rightarrow b\rightarrow 0$), represented by red arrows, corresponding to the direct path for the pair (a, b). (b) A cyclic path ($0\rightarrow a\rightarrow c\rightarrow b\rightarrow 0$) corresponding to the bypass path for the pair (a, b).}
\label{Fig_formulation}
\end{figure}

After adding a dummy node ($i=0$) with edge weights of zero ($w_{k,0}=w_{0,k}=0, {}^\forall k>0$) in the market graph (Fig.~\ref{Fig_formulation}), we seek a cyclic (directed) path giving the minimum weight. Let the node next (/previous) to the dummy node in the cyclic path correspond to the short (/long) positions of a pair trade. As shown in Fig.~\ref{Fig_formulation}, a pair ($a$,~$b$) is represented by both the cyclic path ($0\rightarrow a\rightarrow b\rightarrow 0$) and the cyclic path ($0\rightarrow a\rightarrow c\rightarrow b\rightarrow 0$) with the different weight sums. The former (/latter) representation corresponds to the direct (/bypass) evaluation of the pair ($a$,~$b$). Clockwise and anticlockwise cycles (ex. $0\rightarrow a\rightarrow b\rightarrow 0$ and $0\rightarrow b\rightarrow a\rightarrow 0$) are distinguished.

Define a decision (binary) variable $b_{i,j}$ as taking value 1 if the corresponding edge ($i,j$) is in the chosen cycle and 0 otherwise. The cost function to be minimized is defined by
\begin{equation}\label{Eq_Costfunc}
H_{\mathrm{cost}}=\sum_{i,j}w_{i,j}b_{i,j}.
\end{equation}
The constraints for cyclic directed paths and the tabu search are represented as a penalty function expressed by
\begin{multline}\label{Eq_Penalty}
H_{\mathrm{penalty}}=
\sum_{i}\sum_{j\neq j^{\prime}}b_{i,j}b_{i,j^{\prime}}+
\sum_{j}\sum_{i\neq i^{\prime}}b_{i,j}b_{i^{\prime},j}+ \\
\sum_{i}(\sum_{j}b_{i,j}-\sum_{j}b_{j,i})^2+
\sum_{i,j}b_{i,j}b_{j,i}+
\sum_{i,j}T_{i,j}b_{0,j}b_{i,0}.
\end{multline}
The first (/second) term forces the outflow (/inflow) of each node to be 1 or less. The third term forces the inflows and outflows of each node to be equal. The fourth term forbids traversing the same edge twice in different directions. The fifth term forbids choosing the pairs in the tabu list $T_{i,j}$. Constraint violations increase the penalty, with $H_{\mathrm{penalty}}=0$ if there are no violations. Note that an entry $T_{i,j}$ in the tabu list induces a penalty for the state ($b_{0,j}=b_{i,0}=1$) but not for the states ($b_{0,j}=1$ and $b_{i,0}=0$),  ($b_{0,j}=0$ and $b_{i,0}=1$), and ($b_{i,j}=1$).

The total cost function ($H_{\mathrm{QUBO}}$) is a linear combination of $H_{\mathrm{cost}}$ and $H_{\mathrm{penalty}}$,
\begin{equation}\label{Eq_Cost}
H_{\mathrm{QUBO}}=\sum_{i,j,k,l}Q_{i,j,k,l}b_{i,j}b_{k,l}=m_{c}H_{\mathrm{cost}}+m_{p}H_{\mathrm{penalty}},
\end{equation}
where $m_{c}$ and $m_{p}$ are positive coefficients. The Ising machine searches for the bit configuration $\{b_{i,j}\}$ that minimizes the quadratic cost function $H_{\mathrm{QUBO}}$.

\begin{figure}[t]
\centering
\includegraphics[width=8.3 cm]{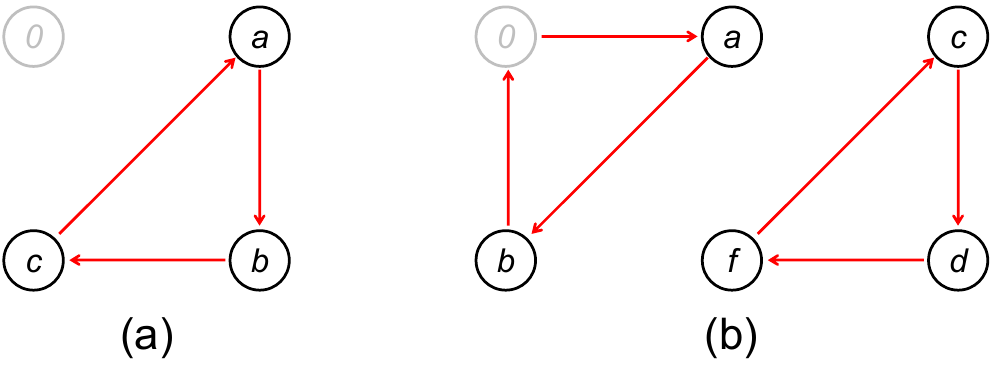}
\caption{Solutions to be excluded by the verification. (a) A cycle without the dummy node. (b) Split cycles.}
\label{Fig_violation}
\end{figure}

The tabu search technique was introduced to enhance the search efficiency upon the multiple executions of the Ising machine for finding multiple opportunities in a market situation under the constraint of forbidding duplicate positions. The procedure and timing of registering and deregistering entries in the tabu list are described in Section~\ref{Sec_System}. In the QUBO formulation, the number of decision variables for an $N$-stock universe is $N(N+1)$ and the size of the solution space (all possible points of the decision variables) is $2^{N(N+1)}$, including constraint violation solutions. We use a heuristic method (an Ising machine) to solve the QUBO problems. Hence, the verification of solutions is necessary and implemented in the system as a function other than Ising machines. In addition, the penalty function, Eq. \eqref{Eq_Penalty}, gives no penalty to the two cases (a cycle without the dummy node and split cycles) shown in Fig.~\ref{Fig_violation}. Those solutions are excluded by the verification. Note that those solutions are not advantageous in the evaluation of the cost function, Eq. \eqref{Eq_Costfunc}.

\section{System}\label{Sec_System} 
To accelerate the decision of opening positions and the issuance of orders after receiving a market feed (informing the change of $ask$ or $bid$ of a stock), the submodules related to the position opening are, in an FPGA, hardwired (instantiated as custom circuits) and inlined as a task pipeline from a receiver (RX) to a transmitter (TX), which are functional without the intervention of a software processor (CPU). The SBM module involved in the pipeline is an inline-type accelerator (not a look-aside type one), featuring a consecutive execution operation and a tabu search function. The management of the positions including the decision of closing positions is carried out by the CPU (software processing). Overall, the system is a hybrid FPGA/CPU system.

\subsection{Architecture}\label{subsec_Archi}
\begin{figure}[t]
\centering
\includegraphics[width=8.3 cm]{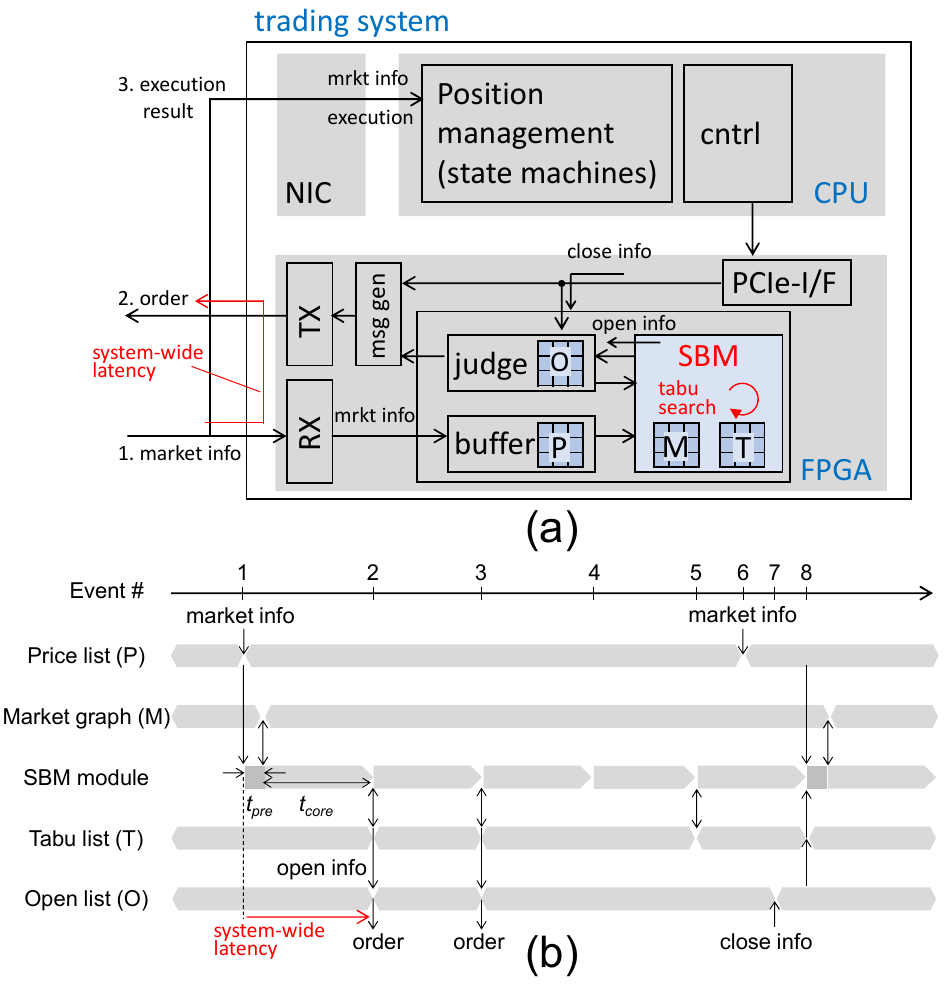}
\caption{System architecture (a hybrid FPGA/CPU system). (a) Block diagram. (b) Timing chart.}
\label{Fig_system}
\end{figure}

Figure~\ref{Fig_system} (a) shows the block diagram of the hybrid FPGA/CPU system. The system components in the FPGA part are, in the order of data flow, a receiver (RX), a price buffer ($P$) that accommodates the price list of $ask$ and $bid$ for the $N$ tradable stocks, the SBM module, a judgment module with a memory unit for the open list ($O$), a message generator, and a transmitter (TX). The SBM module includes two memory units for a market graph ($M$) and a tabu list ($T$), a preprocessing unit ($pre$) for preparing the market graph, and a core processing unit ($core$) for the discrete optimization. Those components are implemented as independent (not synchronized) circuit modules, which are connected with directed streaming data channels with FIFO (first-in-first-out) buffers. The CPU part controls the whole system and manages the positions using state machines for opened positions (see APPENDIX A). The market information (including the changes in $ask$ or $bid$) is received by both the FPGA and CPU parts. The order (buying/selling) packets are issued only from the FPGA part. The execution-result packets informing the results (fill/lapse) of the orders are received by the CPU part. The FPGA and CPU parts are connected with the peripheral component interconnect-express (PCIe) bus.

Figure~\ref{Fig_system} (b) shows the timing chart for the operation of the SBM module when representative events (Events $1$ to $8$) happen. When no event happens for a certain time, the SBM module is idling (polling to the FIFO buffers from the price buffer and judgment modules). When a market feed arrives (Event $1$), the SBM module immediately starts the preprocessing.  The preprocessing unit receives the $2N$ data of $ask$ and $bid$ and then generates the $N(N-1)$ data of weight $w_{i,j}$ (market graph, $M$) with referring to a memory unit for similarity $s_{i,j}$ which is updated once a day before the trading hours. Afterward, the SBM module starts the main (core) processing (the optimal path analysis). Then the SBM module verifies the solution (the path found) in terms of the constraint violations (including the cases of Fig.~\ref{Fig_violation}) and compares the evaluation of the path found with the threshold. If the verification and evaluation pass, the SBM module registers the pair in the tabu list $T$ and concurrently informs it as an open candidate to the judgment module (Event $2$). The judgment module determines the open positions by finally checking them in terms of $P_\mathrm{max}$ (the maximum number of intraday positions) and other control signals, then registers them in the open list $O$ and issues order packets via the message generator (Event $2$).

Here, the judgment module registers the open pair position in the open list $O$ when the opening is decided (before the issuance of orders) and deregisters them when the closing of the pair position is confirmed with the message from the CPU part. When the number of pair positions is decreased, the judgment module informs the updated open list $O$ to the SBM module, which forces the SBM module to refresh the tabu list $T$ by copying the open list $O$ for avoiding duplicate positions.

At the timing of Event $2$, the SBM module starts the main processing again (the consecutive execution operation) without refreshing the tabu list (already up-to-date) and preprocessing (no new market feed arrives), resulting in another order at the timing of Event $3$ (the SBM module could find another tradable path efficiently due to the tabu list). When the SBM does not output an effective solution (Event $4$), the tabu list $T$ and the open list $O$ are not updated. Note that considering the pair based on a direct path (/ a bypass path) corresponding to the ineffective solution may satisfy the threshold if it is evaluated on a bypass path (/ a direct path), we designed that in this case (Event $4$) the pair is not registered in the tabu list. When the SBM outputs an effective solution but it is rejected by the judgment module [for example, due to excess positions ($>P_{\rm max}$)] (Event $5$), the tabu list $T$ is updated but the open list $O$ is not updated. When a new market feed (Event $6$) (or a close confirmation information, Event $7$) arrives, the market graph $M$ (or the tabu list $T$) is updated by the preprocessor (or by copying the open list), at the beginning of the next execution of the SBM module (Event $8$).

As seen in Event $5$, the SBM module determines registering in the tabu list without considering the decision by the judgment module. This design contributes to reducing the latency (not to incorporate the feedback latency from the judgment module). Note that the registration in the tabu list in the case of Event $5$ seems to be undesirable (might miss a trading opportunity) but the over-registration in the tabu list does not matter practically because the tabu list is updated when the positions decrease (Event $8$).

\subsection{Customized SBM core circuit}\label{subsec_Core}
The core processing unit ($core$) is architecturally similar to the basic SBM circuit design~\cite{FPL19} but partially modified for the specific QUBO problem described in Sec.~\ref{subsec_formulation}. The weight $w_{i,j}$ in Eq.~\eqref{Eq_Costfunc} and tabu list $T_{i,j}$ in Eq.~\eqref{Eq_Penalty} are stored in separate memory units (the $M$ memory and $T$ memory in Fig.~\ref{Fig_system}), which are directly accessed by the SBM computation units. Based on the specific pattern of the coupling matrix $Q$, inefficient parts (the products with zero) in the pairwise interaction computation in the SB algorithm are omitted.

In the consecutive execution operation, the SBM module repeats the main processing (simulating the time-evolution of a coupled oscillator network) with different initial states generated by an internal random number generator (RNG), Xorshift RNG~\cite{marsaglia03}. This contributes to efficiently finding another good solution even when the market graph $M$ and the tabu list $T$ are not updated (Event $4$). The latency of the RNG is hidden by overlapping the operations of the SBM core and the RNG; the RNG generates an initial state for the next execution of the SBM core while the SBM core is processing.

\begin{figure}[t]
\centering
\includegraphics[width=8.3 cm]{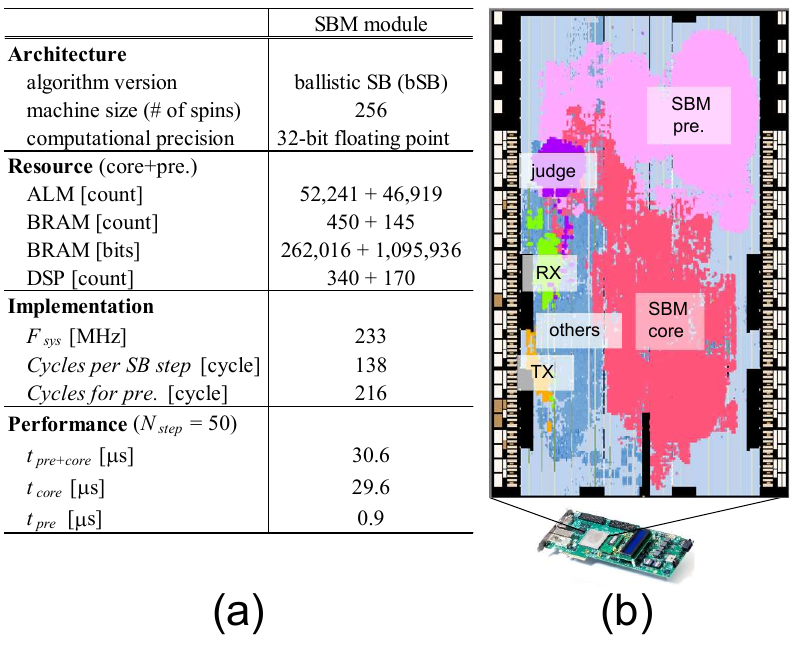}
\caption{System implementation. (a) Architecture and implementation details of the SBM. (b) Placement of functional modules in the FPGA.}
\label{Fig_implementation}
\end{figure}

\subsection{Implementation}
We implemented the system described in Sec.~\ref{subsec_Archi} with a CPU server with a network interface card (NIC) and an FPGA board having another network interface (see APPENDIX B for details).

Figure~\ref{Fig_implementation} (a) shows the architecture and implementation results of the SBM module for 15-stock universes [$N$=15 stocks, $N(N-1)$=210 pairs]. The numbers of nodes and edges (directed) in the market graphs supported are, respectively, 16 and 240, including the dummy node explained in Sec.~\ref{subsec_formulation}. Among three variants of simulated bifurcation (adiabatic, ballistic, and discrete SBs)~\cite{sbm2}, ballistic SB is adopted in this work, with the SB parameters of $N_{\mathrm{step}}$=50 and $dt$=0.65. The machine size (the number of spins) is 256 spins with a specific spin-spin connectively for the QUBO problem described in Sec.~\ref{subsec_formulation}, and the computation precision is 32-bit floating point. Figure~\ref{Fig_implementation} (b) shows the result of the placement of system modules in the FPGA. The SBM module ($core$ and $pre$) is dominant, and the circuit resources used are listed in Fig.~\ref{Fig_implementation} (a). The system clock frequency determined as a result of circuit synthesis, placement, and routing is 233 MHz. The clock cycles of the SB main processing ($core$) and preprocessing ($pre$) are 6,900 steps per run (138 per SB step) and 216, respectively. The computation time (the module latency) per run ($t_{pre}+t_{core}$) is 30.6 $\mu$s, where the SBM core processing is dominant ($t_{core}$=29.6 $\mu$s). The system-wide latency from the market feed arrival to the order packet issuance depicted in Fig.~\ref{Fig_system}(b) as a red arrow is 33 $\mu$s (including the latencies of the RX, price buffer, judgment, SBM, message generator, and TX modules).

\section{Experiment}\label{sec_expt}
The trading system described in Sec.~\ref{Sec_System} was installed at the JPX Co-location area of the TSE and operated through real-time trading to examine whether the strategy based on the consecutive optimal path searches in the $N$-node market graph in Sec.~\ref{sec_strategy} is executable. The trading results are compared with a backcast simulation of the strategy assuming the orders issued are necessarily filled.

The proposed strategy determines the opening of positions based on an instantaneous market situation (a price list of $ask$ and $bid$ for the $N$-stock universe). Because of the latency of a system that executes the strategy and the activities of other trading entities, the orders issued are not necessarily filled at the $ask$/$bid$ prices used for the decision-making. We developed a simulator that processes the historical market feeds provided by the TSE and emulates the internal state of the trading system. The simulator assumes that the orders issued are necessarily filled at the intended prices.

\begin{figure}[t]
\centering
\includegraphics[width=8.3 cm]{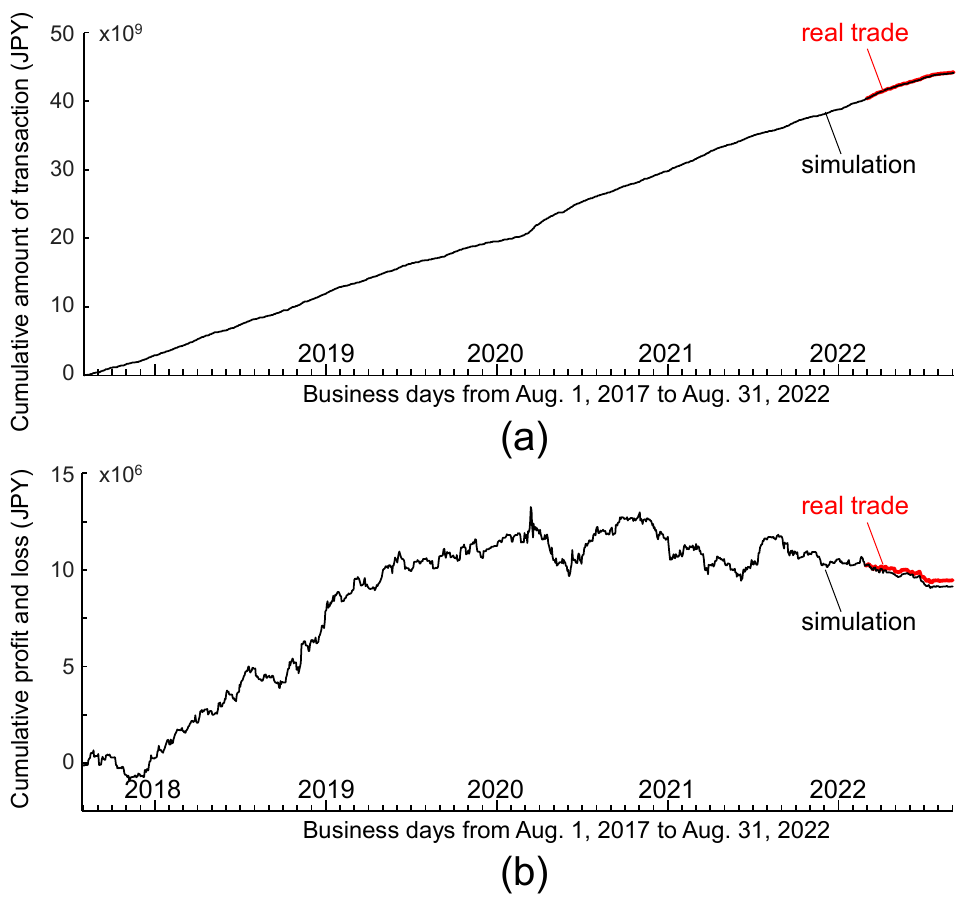}
\caption{Performance of the strategy. (a) Cumulative amount of transaction in billion JPY and (b) cumulative profit and loss in million JPY. Simulation data is from Aug. 1, 2017, to Aug. 31, 2022 (1,239 business days). Real trade data is from Mar. 1, 2022, to Aug. 31, 2022 (125 business days), adjusted with the simulation at the first day.}
\label{Fig_performance}
\end{figure}

Figures \ref{Fig_performance} (a) and (b) show the cumulative values of the amounts of transactions per day and the profit and loss (including $ask$-$bid$ spread costs and commission) per day for real-time trading (red line) and backcast simulation (black line) with fixed strategic parameters of $N$=15 (210 pairs), $P_\mathrm{max}$=16, and $A_\mathrm{trans}$=1.5 million Japanese yen (JPY). The 15 stocks were selected from the bank/insurance sections in terms of high liquidity. The simulation data is from Aug. 1, 2017, to Aug. 31, 2022. The real trade data is from Mar. 1, 2022, to Aug. 31, 2022, being adjusted with the simulation at the first day.

The annualized return and risk over the simulation period (approximately 5 years) are, respectively, 7.5 \% and 9.5 \% for an investment of 24 million JPY  ($A_\mathrm{trans} \times P_{\rm max}$). The Sharpe ratio of the strategy is 0.79, where the Sharpe ratio~\cite{sharpe66} is, in this work, the ratio of the mean to the standard deviation of the return (the profit and loss per period for an investment) from a strategy as in~\cite{backus93}. The strategy proposed can be profitable (a positive annualized return) for the long term (approx. 5 years), especially has shown a high annualized return of 18.5 \% for the period of Aug. 1, 2017, to Feb. 28, 2020, before the COVID-19 pandemic.

The cumulative value of the amount of transaction by the system (3,817,201,458 JPY) over the experiment (750 hours of real-time trading) is coincident well (+2.6 \%) with the simulation value (3,719,389,258 JPY). The fill rate at the intended prices was 93.4 \% and the remaining included the fills at less-favorable prices and the lapses. Most of the lapses occurred just after the opening of the morning sessions. In this experiment, when the order for one of the paired stocks lapses, the position for the other (if the order is filled) is also closed immediately for experimental simplicity (see APPENDIX A), allowing the system to execute more transactions under the constrain of the maximum number of positions ($P_\mathrm{max}$). This is the reason for the increased transaction amount observed in the experiment. Based on the good agreement in the cumulative transaction amounts and detailed comparison analysis of transactions between the experiment and simulation, we conclude that the strategy proposed is executable with the trading system with a latency of 33 $\mu$s.

\begin{figure}[h]
\centering
\includegraphics[width=8.3 cm]{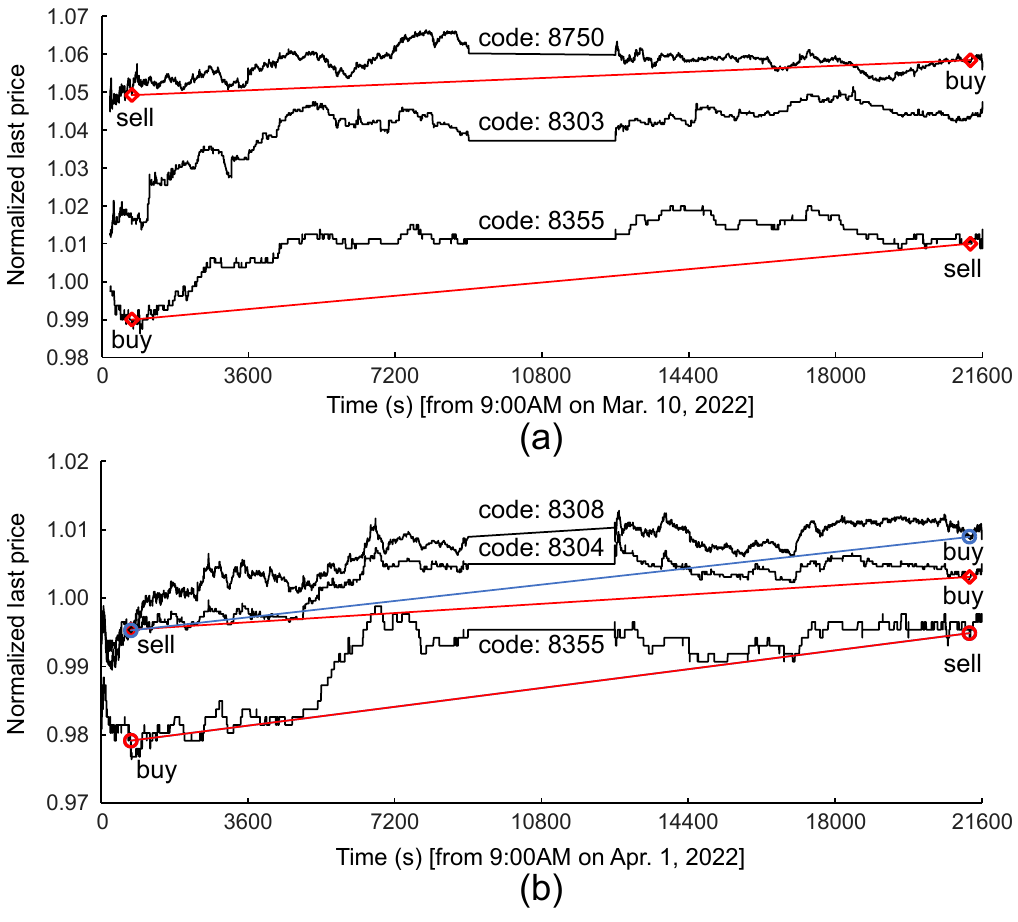}
\caption{Typical transaction behaviors of the trading system on (a) Mar 10, 2022, and (b) Apr. 1, 2022. (a) The open decision (8750, 8355) was made based on the evaluation of the bypass path (8750$\rightarrow$ 8303$\rightarrow$ 8355). (b) Multiple pairs (8308, 8355) and (8304, 8355) were opened in a market situation.}
\label{Fig_typical}
\end{figure}

Figure \ref{Fig_typical} (a) and (b) show typical transaction behaviors by the trading system observed on Mar. 10, 2022, and Apr. 1, 2022, respectively. The number of the market feeds informing the changes of $ask$/$bid$ of stocks in the $N(=15)$-stock universe on Mar. 10 (/Apr. 1) were 1,101,741 (/1,007,773), which arrived at intervals of 18.0 ms (/19.6 ms) on average.

On Mar. 10, 2022, the system decided the opening of the pair position (8750, 8355) [selling code 8750, buying code 8355] at 9:12:14 AM in JST (734 seconds after 9:00:00 AM) based on the evaluation of the bypass path ($8750 \rightarrow 8303 \rightarrow 8355$) found by the SBM module. It was confirmed by the backcast simulation that the evaluation value of the direct path ($8750\rightarrow 8355$) did not satisfy the threshold, meaning that this trading opportunity was missed if the bypass path was not evaluated for decision-making. On that day, both the prices of codes 8750 and 8355 were moving up (uptrend), but the relative difference of the prices (the spread) of the pair position decreased after the position opening, resulting in the profitable closing of the pair position before the end of the trading hours [Fig. \ref{Fig_typical} (a)].

On Apr. 1, 2022, the system decided the opening of the pair positions (8304, 8355) [selling code 8304, buying code 8355] and (8308, 8355) [selling code 8308 buying code 8355] at 9:12:11 AM in JST (731 seconds after 9:00:00 AM) based on the evaluation of the direct paths ($8304 \rightarrow 8355$) and ($8308 \rightarrow 8355$). The two pair positions were found by the consecutive execution operation of the SBM module in the instantaneous market situation (before the market situation changed). On that day, the prices of codes 8308, 8304, and 8355 were, overall, moving up (uptrend), but the spreads of the pair positions decreased after the position opening, resulting in the profitable closing of the pair positions before the end of the trading hours [Fig. \ref{Fig_typical} (b)].

\section{Conclusion}
We proposed a pairs-trading strategy that finds trading opportunities for any two stocks in an $N$-stock universe through solving an optimal path search problem in market graphs and have demonstrated with the real-time transaction records in the TSE that the strategy is executable in terms of response latency with the automated trading system using the SB-based embeddable Ising machine for the market graph analysis.

The market graph for the $N$-stock universe is an $N$-node fully-connected directed graph with each edge weight corresponding to the product of instantaneous price difference and dynamic time warping (DTW) distance-based similarity between a pair of stocks. In the graph, two nodes connected by the minimum-weight one-way directed path selected from among all possible direct and bypass paths (a collective evaluation of the graph) are considered to correspond to the best trading opportunity. The optimal path search is consecutively executed to find multiple trading opportunities in an instantaneous market situation with avoiding duplicate detections by a tabu search technique.

The automated trading system is a hybrid FPGA/CPU system. The FPGA part (hardware processing) decides the opening of a pair of long/short positions using the Ising machine and then issues the corresponding orders, while the CPU part (software processing) manages the opened positions (including the decision of closing positions). The system-wide latency from the market feed arrival to the order packet issuance is 33 $\mu$s for $N$=15 or 210 pairs.

The trading system was installed at the JPX Co-location area of the TSE and operated for a real-time trading period of 125 business days or 750 hours. The real-time transaction records were compared with a backcast simulation of the strategy assuming the orders issued are necessarily filled at the intended prices. Based on the good agreement in the cumulative transaction amounts and detailed comparison analysis of transactions between the experiment and simulation, we have concluded that the response latency of the system with the SB-based Ising machine is sufficiently low to execute the pairs-trading strategy based on optimal path search in market graphs.

Automated trading systems with embedded Ising machines would be applicable to the strategies based on various graph analyses of market graphs defined by various return/risk measures and other trading strategies that rely on high-speed discrete optimization.

\section*{Appendices}
\subsection*{A. Position management}
\begin{figure}[h]
\centering
\includegraphics[width=4.0 cm]{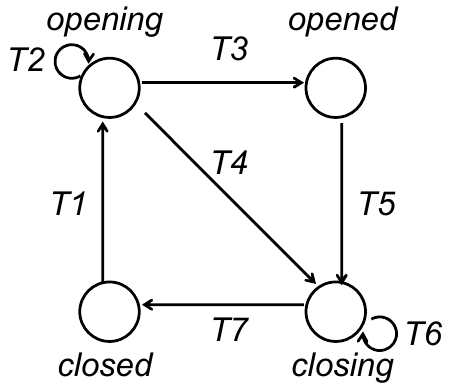}
\caption{State diagram for a state machine for the pair position management.}
\label{Fig_state_machine}
\end{figure}

The position management module manages $N(N-1)$ state machines corresponding to all the pairs. Fig.~\ref{Fig_state_machine} shows the states and transitions of the state machine. Initially the pair position ($i$,~$j$) has been closed ($closed$ state). When an execution packet (informing that the order of one of the stock pair is filled) is received, the state shifts to $opening$ state ($T1$ transition) and then stays waiting for the remaining results to be received ($T2$). If the fill of the orders for the pair is confirmed as intended, the state shifts to $opened$ ($T3$). Otherwise (unintended), the state shifts to $closing$ ($T4$). The management module always monitors the prices ($bid$ and $ask$) of all the tradable stocks and detects the convergence of the spread when opened (the confirmation of a profit more than a threshold) for the opened pair. If the closing condition is satisfied, the state shifts to $closing$ ($T5$). In the $closing$ state, the state stays waiting for the related positions to be all closed ($T6$); the management module issues the orders for closing via the message generator in the FPGA and then (if necessary) repeats ordering until all the positions are closed. If the closing of the positions is confirmed, the state shifts back to $closed$ ($T7$).

\subsection*{B. Implementation details}
An FPGA board and a high-speed network interface card (NIC) are mounted on a host server with dual CPUs (Intel Xeon Silver 4215R) and DDR-DRAM modules (384 GB). The FPGA (Intel Arria 10 GX 1150 FPGA) on the board has 427,200 adaptive logic modules (ALMs) including 854,400 adaptive look-up-tables (ALUTs, 5-input LUT equivalent) and 1,708,800 flip-flop registers, 2,713 20Kbit-size RAM blocks (BRAMs), and 1,518 digital signal processor blocks (DSPs). The system components in the FPGA described in Section~\ref{Sec_System} were coded in a high-level synthesis (HLS) language (Intel FPGA SDK for OpenCL, ver.~18.1). The FPGA interfaces including a PCIe IP (PCIe Gen3$\times$8), a 10~Gbps Ethernet PHY IP and communication IPs (RX, TX) were written in Verilog HDL and incorporated in the board support package (BSP).

\subsection*{Acknowledgment}
The experiment in the Tokyo Stock Exchange was conducted under a joint project between Toshiba Corporation and Dharma Capital. K.K. The authors thank Ryosuke Iio and Kohei Shimane for fruitful discussions and technical support.

\subsection*{Conflicts of Interest}
K.T., R.H., and M.Y. are included in inventors on two U.S. patent applications related to this work filed by the Toshiba Corporation (no. 17/249353, filed 20 February 2020; no. 17/565206, filed 29 December 2021). The authors declare that they have no other competing interests.

\begin{figure}[h]
\vspace{0.5cm}
\noindent\includegraphics[width=1in,height=1.25in,clip,keepaspectratio]{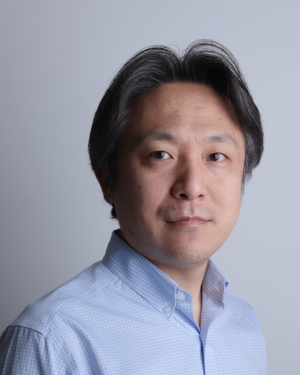}\\
{\small Kosuke Tatsumura received the B.E., M.E., and Ph.D. degrees in Electronics, Information and Communications Engineering from Waseda University, Japan, in 2000, 2001, and 2004, respectively. After working as a postdoctoral fellow at Waseda University, he joined Toshiba Corporation in 2006. He is a chief research scientist, leading a research team and several projects toward realizing innovative industrial systems based on cutting-edge computing technology. He was a member of the Emerging Research Devices (ERD) committee in the International Technology Roadmap for Semiconductors (ITRS) from 2013 to 2015. He has been a lecturer at Waseda University since 2013. He was a visiting researcher at the University of Toronto from 2015 to 2016. He received the Best Paper Award at IEEE Int. Conf. on Field-Programmable Technology (FTP) in 2016. His research interests include domain-specific computing, quantum/quantum-inspired computing, and their applications.}
\vspace{-0.5cm} 
\end{figure}

\begin{figure}[h]
\vspace{0.5cm}
\noindent\includegraphics[width=1in,height=1.25in,clip,keepaspectratio]{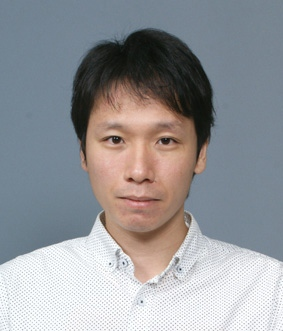}\\
{\small Ryo Hidaka received the B.E. and M.E. degrees in Systems Design and Informatics from Kyushu Institute of Technology, Japan, in 2006 and 2008, respectively. He joined Toshiba Corporation in 2008. He was engaged in the development of main processors (2D-to-3D conversion and local dimming) for digital televisions, an image recognition processor called Visconti\textsuperscript{TM}, and host controllers for flash-memory cards. His current research interests include domain-specific computing, high-level synthesis design methodology, and proof-of-concept study with FPGA devices.}
\vspace{-0.5cm} 
\end{figure}

\begin{figure}[h]
\vspace{0.5cm}
\noindent\includegraphics[width=1in,height=1.25in,clip,keepaspectratio]{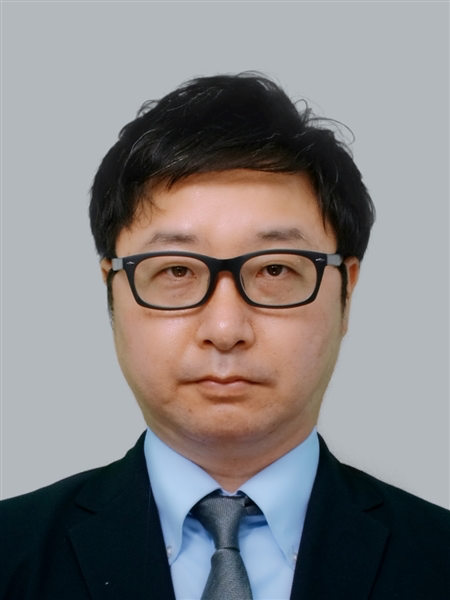}\\
{\small Jun Nakayama received the degree of Bachelor of Arts in Economic and Social studies from the University of Manchester, the U.K., in 2008. He received the degree of Master of Business Administration (MBA) in Finance from Hitotsubashi University, Japan, in 2017. He was a portfolio manager in Nomura Asset Management Co., Ltd. from 2008 to 2020 and was engaged in the development and management of quant-based funds. He joined Toshiba Corporation in 2020. He is also a Ph.D. candidate in the Financial Strategy Program, Hitotsubashi University Business School. His research interests include quantitative investment strategies, quantum-inspired computing technology, and trading strategies with advanced technologies.}
\vspace{-0.5cm} 
\end{figure}

\begin{figure}[h]
\vspace{0.5cm}
\noindent\includegraphics[width=1in,height=1.25in,clip,keepaspectratio]{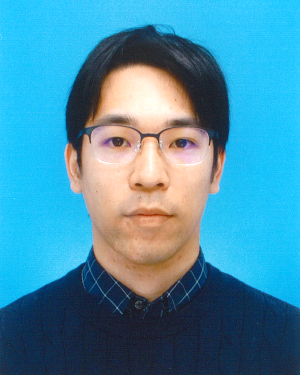}\\
{\small Tomoya Kashimata received the B.E., and M.E. degrees in computer science and engineering from Waseda University, Japan, in 2018 and 2020, respectively. He joined Corporate Research and Development Center, Toshiba Corporation, Japan, in 2020. His research interests include computer architecture, reconfigurable architecture, and processor in memory.}
\vspace{-0.5cm} 
\end{figure}

\begin{figure}[h]
\vspace{0.5cm}
\noindent\includegraphics[width=1in,height=1.25in,clip,keepaspectratio]{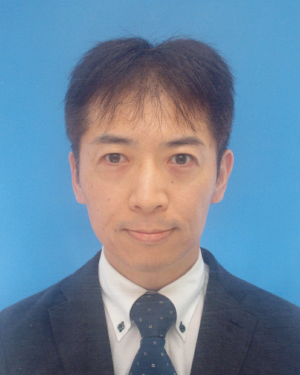}\\
{\small Masaya Yamasaki received the B.E. and M.E. degrees in Computer Science and Communication Engineering from Kyushu University, Japan, in 1997 and 1999, respectively. He joined Toshiba Corporation in 1999. He was engaged in the development of image processing engines (interframe interpolation technology) for digital televisions (including ones with Cell Broadband Engines) and FPGA-based coprocessors for multi-channel video recording, three-dimensional display, and industrial systems. His research interests include domain-specific computing, high-level synthesis design space exploration, and proof-of-concept study with FPGA devices.}
\vspace{-0.5cm} 
\end{figure}


\begin{thebibliography}{00}
{\small 
\bibitem{Sharpe90} W. F. Sharpe, G. J. Alexander, J. V. Bailey, ``Investments (4th edition),'' Prentice Hall, Englewood Cliffs, N.J.  1990.

\bibitem{shleifer97} A. Shleifer, R. W. Vishny, ``The limits of arbitrage,'' \emph{The Journal of finance} \textbf{52}, pp. 35--55, 1997. [Online]. Available: https://doi.org/10.1111/j.1540-6261.1997.tb03807.x

\bibitem{gromb10}
D. Gromb, D. Vayanos, ``Limits of arbitrage,'' \emph{Annual Review of Financial Economics} \textbf{2}, pp. 251--275, 2010. [Online]. Available: https://doi.org/10.1146/annurev-financial-073009-104107

\bibitem{rosch21}
D. R{\"o}sch, ``The impact of arbitrage on market liquidity,'' \emph{Journal of Financial Economics} \textbf{142}, pp. 195--213, 2021. [Online]. Available: https://doi.org/10.1016/j.jfineco.2021.04.034

\bibitem{gatev06} E. Gatev, W. N. Goetzmann, K. G. Rouwenhorst, ``Pairs trading: Performance of a relative-value arbitrage rule,'' \emph{The Review of Financial Studies} \textbf{19}, pp. 797--827, 2006. [Online]. Available: https://doi.org/10.1093/rfs/hhj020

\bibitem{krauss17} C. Krauss, ``Statistical arbitrage pairs trading strategies: Review and outlook,'' \emph{Journal of Economic Surveys} \textbf{31}, pp. 513--545, 2017. [Online]. Available: https://doi.org/10.1111/joes.12153

\bibitem{flori21} A. Flori, D. Regoli, ``Revealing pairs-trading opportunities with long short-term memory network,'' \emph{European Journal of Operational Research} \textbf{295}, pp. 772--791, 2021. [Online]. Available: https://doi.org/10.1016/j.ejor.2021.03.009

\bibitem{butenko03} S. Butenko, ``Maximum independent set and related problems, with applications,'' \emph{Ph.D. dissertation, the Industrial and Systems Engineering Department, University of Florida}, 2003. [Online]. Available: https://ufdcimages.uflib.ufl.edu/UF/E0/00/10/11/\\00001/butenko\_s.pdf

\bibitem{boginski04} V. Boginski,  S. Butenko, P. M. Pardalos,  ``Network-based Techniques in the Analysis of the Stock Market,'' in \emph{Supply Chain and Finance}, eds. P. M. Pardalos, A. Migdalas, G. Baourakis, World Scientific, pp. 1--14, 2004. [Online]. Available: https://doi.org/10.1142/9789812562586\_0001

\bibitem{marzec16} M. Marzec, ``Portfolio optimization: Applications in quantum computing,'' in \emph{Handbook of High-Frequency Trading and Modeling in Finance} eds. I. Florescu, M. C. Mariani, H. E. Stanley, F. G. Viens, Wiley Online Library, pp. 73--106, 2016. [Online]. Available: https://doi.org/10.1002/9781118593486.ch4

\bibitem{lucas14} A. Lucas, ``Ising formulations of many NP problems,'' \emph{Frontiers in physics} \textbf{2}, 5, 2014. [Online]. Available: https://doi.org/10.3389/fphy.2014.00005

\bibitem{sbm1} H. Goto, K. Tatsumura, A. R. Dixon, ``Combinatorial optimization by simulating adiabatic bifurcations in nonlinear Hamiltonian systems,'' \emph{Science Advances} \textbf{5}, eaav2372, 2019. [Online]. Available: https://doi.org/10.1126/sciadv.aav2372

\bibitem{FPL19} K. Tatsumura, A. R. Dixon, H. Goto, ``FPGA-Based Simulated Bifurcation Machine,'' \emph{Proc. of IEEE International Conference on Field Programmable Logic and Applications} (FPL), pp. 59--66, 2019. [Online]. Available: https://doi.org/10.1109/FPL.2019.00019

\bibitem{sbm2} H. Goto, K. Endo, M. Suzuki, Y. Sakai, T. Kanao, Y. Hamakawa, R. Hidaka, M. Yamasaki, K. Tatsumura, ``High-performance combinatorial optimization based on classical mechanics,'' \emph{Science Advances} \textbf{7}, eabe7953, 2021. [Online]. Available: https://doi.org//10.1126/sciadv.abe7953

\bibitem{NatEle} K. Tatsumura, M. Yamasaki, H. Goto, ``Scaling out Ising machines using a multi-chip architecture for simulated bifurcation,'' \emph{Nature Electronics} \textbf{4}, pp. 208--217, 2021. [Online]. Available: https://doi.org/10.1038/s41928-021-00546-4

\bibitem{kanao23} T. Kanao, H. Goto, ``Simulated bifurcation for higher-order cost functions,'' \emph{Applied Physics Express} \textbf{16}, 014501, 2023. [Online]. Available: https://doi.org/10.35848/1882-0786/acaba9

\bibitem{johnson11}
M. W. Johnson, M. H. S. Amin, S. Gildert, T. Lanting, F. Hamze, N. Dickson, R. Harris, A. J. Berkley, J. Johansson, P. Bunyk, E. M. Chapple, C. Enderud, J. P. Hilton, K. Karimi, E. Ladizinsky, N. Ladizinsky, T. Oh, I. Perminov, C. Rich, M. C. Thom, E. Tolkacheva, C. J. S. Truncik, S. Uchaikin, J. Wang, B. Wilson, G. Rose, ``Quantum annealing with manufactured spins,'' \emph{Nature} \textbf{473}, pp. 194--198 (2011). [Online]. Available: https://doi.org/10.1038/nature10012

\bibitem{king23}
A. D. King, J. Raymond, T. Lanting, R. Harris, A. Zucca, F. Altomare, A. J. Berkley, K. Boothby, S. Ejtemaee, C. Enderud, E. Hoskinson, S. Huang, E. Ladizinsky, A. J. R. MacDonald, G. Marsden, R. Molavi, T. Oh, G. Poulin-Lamarre, M. Reis, C. Rich, Y. Sato, N. Tsai, M. Volkmann, J. D. Whittaker, J. Yao, A. W. Sandvik, M. H. Amin, ``Quantum critical dynamics in a 5,000-qubit programmable spin glass,'' \emph{Nature} \textbf{617}, pp. 61–-66 (2023). [Online]. Available: https://doi.org/10.1038/s41586-023-05867-2

\bibitem{honjo21}
T. Honjo, T. Sonobe, K. Inaba, T. Inagaki, T. Ikuta, Y. Yamada, T. Kazama, K. Enbutsu, T. Umeki, R. Kasahara, K. Kawarabayashi, H. Takesue, ``100,000-spin coherent ising machine,'' \emph{Science Advances} \textbf{7}, eabh095 (2021). [Online]. Available: https://doi.org/10.1126/sciadv.abh0952

\bibitem{pierangeli19}
D. Pierangeli, G. Marcucci, C. Conti, ``Large-Scale Photonic Ising Machine by Spatial Light Modulation,'' \emph{ Physical Review Letters} \textbf{122}, 213902 (2019). [Online]. Available: https://doi.org/10.1103/PhysRevLett.122.213902

\bibitem{cai20} F. Cai, S. Kumar, T. V. Vaerenbergh, X. Sheng, R. Liu, C. Li, Z. Liu, M. Foltin, S. Yu, Q. Xia, J. J. Yang, R. Beausoleil, W. D. Lu, J. P. Strachan, ``Power-efficient combinatorial optimization using intrinsic noise in memristor Hopfield neural networks,'' \emph{Nature Electronics} \textbf{3}, pp. 409--418, 2020. [Online]. Available: https://doi.org/10.1038/s41928-020-0436-6

\bibitem{aadit22} N. A. Aadit, A. Grimaldi, M. Carpentieri, L. Theogarajan, J. M. Martinis, G. Finocchio, K. Camsari, ``Massively parallel probabilistic computing with sparse Ising machines,'' \emph{Nature Electronics} \textbf{5}, pp. 460--468, 2022. [Online]. Available: https://doi.org/10.1038/s41928-022-00774-2

\bibitem{moy22} W. Moy, I. Ahmed, P. Chiu, J. Moy, S. S. Sapatnekar, C. H. Kim, ``A 1,968-node coupled ring oscillator circuit for combinatorial optimization problem solving,'' \emph{Nature Electronics} \textbf{5}, pp. 310--317, 2022. [Online]. Available: https://doi.org/10.1038/s41928-022-00749-3

\bibitem{sharma22} A. Sharma, R. Afoakwa, Z. Ignjatovic, M. Huang, ``Increasing Ising machine capacity with multi-chip architectures,'' \emph{Proc. of Annual International Symposium on Computer Architecture} (ISCA), pp. 508--521, 2022. [Online]. Available: https://doi.org/10.1145/3470496.3527414

\bibitem{takemoto19}
T. Takemoto, M. Hayashi, C. Yoshimura, M. Yamaoka, ``A 2$\times$30k-Spin Multi-Chip Scalable Annealing Processor Based on a Processing-In-Memory Approach for Solving Large-Scale Combinatorial Optimization Problems,'' \emph{IEEE Journal of Solid-State Circuits} \textbf{55}, pp. 145--156, 2019. [Online]. Available: https://doi.org/10.1109/JSSC.2019.2949230

\bibitem{kawamura23} K. Kawamura, J. Yu, D. Okonogi, S. Jimbo, G. Inoue, A. Hyodo, {\'A}. L. Garc{\'\i}a-Anas, K. Ando, B. H. Fukushima-Kimura, R. Yasudo, T. Van Chu, M. Motomura, ``Amorphica: 4-replica 512 fully connected spin 336MHz metamorphic annealer with programmable optimization strategy and compressed-spin-transfer multi-chip extension,'' {Proc. of  IEEE International Solid-State Circuits Conference} (ISSCC), pp. 42--43, 2023. [Online]. Available: https://doi.org/10.1109/ISSCC42615.2023.10067504

\bibitem{matsubara20} S. Matsubara, M. Takatsu, T. Miyazawa, T. Shibasaki, Y. Watanabe, K. Takemoto, H. Tamura, ``Digital annealer for high-speed solving of combinatorial optimization problems and its applications,'' \emph{Proc. of Asia and South Pacific Design Automation Conference} (ASP-DAC), pp. 667--672, 2020. [Online]. Available: https://doi.org/10.1109/ASP-DAC47756.2020.9045100

\bibitem{waidyasooriya21} H. M. Waidyasooriya, M. Hariyama, ``Highly-parallel FPGA accelerator for simulated quantum annealing,'' \emph{ IEEE Transactions on Emerging Topics in Computing} \textbf{9}, pp. 2019–2029, 2021. [Online]. Available: https://doi.org/10.1109/TETC.2019.2957177

\bibitem{okuyama19} T. Okuyama, T. Sonobe, K. Kawarabayashi, M. Yamaoka, ``Binary optimization by momentum annealing,'' \emph{ Physical Review E} \textbf{100}, 012111, 2019. [Online]. Available: https://doi.org/10.1103/PhysRevE.100.012111

\bibitem{barahona82} F. Barahona, ``On the computational complexity of Ising spin glass models,'' \emph{Journal of Physics A: Mathematical and General} \textbf{15}, pp. 3241–-3253, 1982. [Online]. Available: https://doi.org/10.1088/0305-4470/15/10/028

\bibitem{yoo23} S. Yoo, H. Kim, J. Kim, S. Park, J.-Y. Kim, J. Oh, ``LightTrader: A Standalone High-Frequency Trading System with Deep Learning Inference Accelerators and Proactive Scheduler,'' \emph{IEEE International Symposium on High-Performance Computer Architecture} (HPCA), pp. 1017--1030, 2023. [Online]. Available: https://doi.org/10.1109/HPCA56546.2023.10070930

\bibitem{fil20} M. Fil, L. Kristoufek, ``Pairs trading in cryptocurrency markets,'' \emph{IEEE Access} \textbf{8}, pp. 172644--172651, 2020. [Online]. Available: https://doi.org/10.1109/ACCESS.2020.3024619

\bibitem{huang19} B. Huang, Y. Huan, L. D. Xu, L. Zheng, Z. Zou, ``Automated trading systems statistical and machine learning methods and hardware implementation: a survey,'' \emph{Enterprise Information Systems} \textbf{13}, pp. 132--144, 2019. [Online]. Available: https://doi.org/10.1080/17517575.2018.1493145

\bibitem{denholm15} S. Denholm, H. Inoue, T. Takenaka, T. Becker, W. Luk, ``Network-level FPGA acceleration of low latency market data feed arbitration,'' \emph{IEICE Transactions on Information and Systemss} \textbf{E98-D}, pp. 288--297, 2015. [Online]. Available: https://doi.org/10.1587/transinf.2014RCP0011

\bibitem{leber11} C. Leber, B. Geib, H. Litz, ``High frequency trading acceleration using FPGAs,'' \emph{Proc. of IEEE International Conference on Field Programmable Logic and Applications} (FPL), pp. 317--322, 2011. [Online]. Available: https://doi.org/10.1109/FPL.2011.64

\bibitem{malceniece23} L. Malceniece, K. Malcenieks, T. J. Putni{\c{n}}{\v{s}}, T{\=a}lis, ``High frequency trading and comovement in financial markets,'' \emph{Journal of Financial Economics} \textbf{134}, pp. 381--399, 2019. [Online]. Available: https://doi.org/10.1016/j.jfineco.2018.02.015

\bibitem{brogaard14} J. Brogaard, T. Hendershott, R. Riordan, ``High-Frequency Trading and Price Discovery,'' \emph{The Review of Financial Studies} \textbf{27}, pp. .2267-2306, 2014. [Online]. Available: https://doi.org/10.1093/rfs/hhu032

\bibitem{spyrou13}  S. Spyrou, ``Herding in financial markets: a review of the literature,'' \emph{Review of Behavioral Finance},\textbf{5}, pp. 175--194, 2013. [Online]. Available: https://doi.org/10.1108/RBF-02-2013-0009

\bibitem{ISCAS20} K. Tatsumura, R. Hidaka, M. Yamasaki, Y. Sakai, H. Goto, ``A Currency Arbitrage Machine based on the Simulated Bifurcation Algorithm for Ultrafast Detection of Optimal Opportunity,'' \emph{Proc. of IEEE International Symposium on Circuits and Systems} (ISCAS), pp. 1--5, 2020. [Online]. Available: https://doi.org/10.1109/ISCAS45731.2020.9181114

\bibitem{qbm} H. Goto, ``Bifurcation-based adiabatic quantum computation with a nonlinear oscillator network,'' \emph{Scientific Reports} \textbf{6}, 21686, 2016. [Online]. Available: https://doi.org/10.1038/srep21686

\bibitem{sakoe78} H. Sakoe, S. Chiba, ``Dynamic programming algorithm optimization for spoken word recognition,'' \emph{IEEE Transactions on Acoustics, Speech, and Signal Processing} \textbf{26}, pp. 43--49, 1978. [Online]. Available: https://doi.org/10.1109/TASSP.1978.1163055

\bibitem{marsaglia03} G. Marsaglia, ``Xorshift RNGs,'' \emph{Journal of Statistical software} \textbf{8}, pp. 1--6, 2003. [Online]. Available: https://doi.org/10.18637/jss.v008.i14

\bibitem{sharpe66} W. F. Sharpe,  ``Mutual fund performance,'' \emph{The Journal of Business} \textbf{39}, pp. 119--138, 1966. [Online]. Available: https://www.jstor.org/stable/2351741

\bibitem{backus93} D. K. Backus, A. W. Gregory, C. I. Telmer, ``Accounting for forward rates in markets for foreign currency,'' \emph{The Journal of Finance} \textbf{48}, pp. 1887--1908, 1993. [Online]. Available: https://doi.org/10.1111/j.1540-6261.1993.tb05132.x
}
\end{thebibliography}
\end{document}